\documentstyle[epsf,12pt]{article}
\begin{document}
\begin{titlepage}
\begin{center}
\vspace*{4cm}

\begin{title}
\bold {\Huge 
Implementation of Bose - Einstein 
interference effects 
in Monte Carlo generators}

\end{title}
\vspace{2cm}

\begin{author}
\Large {K. FIA{\L}KOWSKI\footnote{e-mail address:
uffialko@thrisc.if.uj.edu.pl} and R. WIT }
\end{author}\\
Institute of Physics, Jagellonian University \\
30-059 Krak{\'o}w, ul.Reymonta 4, Poland

\vspace{3cm}
\begin{abstract}
   A practical method of calculating weights implementing the  Bose - Einstein  interference  effects
is presented.  The usefulness of the method is demonstrated for  the events generated by
the JETSET/PYTHIA
code to describe the {\sl UA1} proton-antiproton data at 630GeV. A good description
of data
  is achieved with a reasonable value of the Gaussian width of a two-particle weight
factor, which is the only free parameter in our calculations.
\end{abstract}
\end{center}
\end{titlepage}

\section{Introduction}
\par
Recently various methods of imitating  the  Bose  -
Einstein interference effects in Monte Carlo generators have been 
discussed [1-3]. This problem became acute with the advent of new data on WW 
production, since quite conflicting estimates of the W mass shifts resulting
from these effects have been published [4-5].
\par
It has been shown that the most commonly used method of momenta shifting
(employed in the new versions of JETSET [6]) does not reproduce the input shape
of the Bose - Einstein correlation factor, even after improvements of the
original procedure [3]. Thus modifying this shape to fit the data we cannot hope
to learn much about the space-time evolution of the production process.
\par
On the other hand, the method attributing weights to generated events, much
better
justified theoretically [7],  as recently discussed in more detail by Bia{\l}as and
Krzywicki [8] is not easy to implement. The
factorial increase of the number of terms to be calculated with multiplicity
makes impossible the direct application of the method for high energies. Only
recently Wosiek [9] has indicated a possible scheme shortening such
calculations. Symmetrizing separately the particles from each hemisphere, as
proposed by Haywood [1], introduces the unknown bias against the effect for
slow particles (in the CM). In addition, this procedure does not remove the
fundamental difficulties but only shifts them to  higher
multiplicities. Another problem of the weight method is the serious distorsion of the multiplicity
distribution, since weights enhance the high multiplicity tail [2].
\par
Solution to these problems has been proposed recently by Jadach and Zalewski [5], 
who reduce the number of terms replacing the original Bia{\l}as - Krzywicki 
formula by an approximation based on the clustering algorithm. The initial
 average multiplicity is restored by rescaling the weights with a simple $cV^n$
factor.
\par
In this note we propose another approximation which seems to be quickly
convergent to the full formula without prohibitive increase in the computing
time. We present the method in the next section and then apply it to the
description of the {\sl UA1} data [10]. The results are encouraging. We hope to use the
same method to clarify the problem of the W mass shifts in hadronic decays.

\section{Calculating weights for multiparticle states}
\par
According to the Bia{\l}as - Krzywicki prescription, the symmetrization of
amplitudes required by the Bose - Einstein statistics may be approximated by
generating unsymmetrized distributions and correcting them {\sl a posteriori} by the
multiplicative weights attributed to each event. Such a weight is a sum over all
permutations of identical particles of the products of two particle weight
factors  $w_{iP(i)}$  calculated for the pair of momenta (of i-th
particle and the particle,
which occupies the i-th place in the permutation $P\{i\}$).
\begin{equation}
\label{s1}
W(n) = \sum_{\{P(i)\}}\prod_{i=1}^{n}w_{iP(i)}.
\end{equation}
 Since all factors are positive and $w_{ii} =1$,
the resulting weight is not smaller than one (a contribution from identity
permutation). One may correct results to keep, e.g., the average number of
particles fixed; we return to this point later. 
\par
Since most of the particles detected in experiments are pions, the final weight
should be actually given by a product of weights calculated separately for
positive, negative and neutral pions.  In fact, the  BE interference for neutral
particles is not observable (apart from the possible effects for hard photons
[11]): neutral pions decay before
detection, and for the resulting photons the effective source size is so big
that the BE effects must be negligible for momentum differences above a few eV.
However, the procedure should not change the observable correlations between the
numbers of charged and neutral pions. Therefore weights for all signs of pions
must be taken into account.
\par
Thus in principle the only arbitrary factor is the function of the difference of
two momenta
$w_{ij}(p_i -  p_j)$. It
is natural to try as a first guess the Gaussian function of four-momentum
difference squared 
\begin{equation}
\label{s2}
w_{ij} = e^{(p_i - p_j)^2/2\sigma ^2}.
\end{equation}  
Of course,  different components of momentum difference squared may be
multiplied by different coefficients, and the shape may be modified. In this
note we do not discuss these possibilities. Therefore the only parameter is a
Gaussian half-width of the distribution $\sigma$.
\par 
We have tried first to implement this prescription generating by the
JETSET/PYTHIA program [6,12] the samples of $10^5$ events
 of $\overline{p} p$ collisions at 10 and 30
GeV CM energies. For each event for each permutation the product of two-particle
weight factors (with $\sigma\ = 0.14$ GeV which corresponds to 1 fm radius of the
Fourier transform of the weight (2)) is computed and all contributions are
added to calculate the weight for the event which is used to produce
distributions (to be compared with those for all weights equal one).
\par
Unfortunately, for more than ten pions of a given sign the calculations become
prohibitively long. This does not happen at 10 GeV, but already at 30 GeV there
are more then hundred events with such multiplicities. To get any results  we
had to exclude them from standard weight calculations (attributing to each of them the
same value of the weight as obtained for the previously generated  event). This removes, however,
the fluctuations which are most interesting for the investigation of very short
range correlations.
\par
Thus here we have separated the sum of all the  n! permutations 
into terms where only the permutations which change places of exactly $K$ particles
are taken into account:
\begin{equation}
\label{s3}
w = \sum_Kw^{(K)}.
\end{equation}

These terms for $K<6$ are

$$
\begin{array}{lll}
w^{(0)} = 1; \   w^{(1)} = 0; \  
w^{(2)} = \sum_{i=1}^{n-1}\sum_{j>i}(w_{ij})^2; \ 
w^{(3)}  = 2\sum_{i=1}^{n-2}\sum_{j>i}\sum_{k>j}w_{ij}w_{jk}w_{ki}; & &
\\
 & & \\
w^{(4)} = 
\sum_{i=1}^{n-3}\sum_{j>i}\sum_{k>j}\sum_{l>k}[2w_{ij}w_{ik}w_{jl}w_{kl}
+2w_{ij}w_{il}w_{jk}w_{kl} +2w_{ik}w_{il}w_{jk}w_{jl} + (w_{il}w_{jk})^2 + & &
\\
 & &  \\ 
(w_{ij}w_{kl})^2 + (w_{ik}w_{jl})^2]; & & 
\\
 & & \\
w^{(5)} =
2\sum_{i=1}^{n-4}\sum_{j>i}\sum_{k>j}\sum_{l>k}\sum_{m>l}[(w_{ij})^2w_{lk}w_{ml}w_{km}
+(w_{ik})^2w_{jl}w_{ml}w_{jm} + &&  
\\
\end{array}
$$
\begin{equation}
\label{s4}
\begin{array}{lll}
(w_{il})^2w_{jk}w_{jm}w_{km} + (w_{im})^2w_{jk}w_{kl}w_{jl} + (w_{jk})^2w_{il}w_{lm}w_{im}
+(w_{jl})^2w_{ik}w_{km}w_{im} +  & &
\\
 & &\\
(w_{jm})^2w_{ik}w_{kl}w_{il} + (w_{kl})^2w_{ij}w_{jm}w_{im} +(w_{lm})^2w_{ij}w_{jk}w_{ik} +
 (w_{km})^2w_{ij}w_{jl}w_{il} +  && \\ 
&& \\
w_{ij}w_{jk}w_{kl}w_{lm}w_{im} +  w_{ik}w_{jl}w_{km}w_{jm}w_{il} +w_{il}w_{ij}w_{kl}w_{jm}w_{km} + 
w_{ij}w_{ik}w_{jl}w_{lm}w_{km} + && \\
&& \\w_{ik}w_{im}w_{jk}w_{jl}w_{lm} +  
w_{il}w_{jl}w_{jk}w_{km}w_{im} + w_{ij}w_{ik}w_{kl}w_{lm}w_{jm} + 
w_{ij}w_{il}w_{lm}w_{jk}w_{km} + && \\
&& \\
w_{ij}w_{im}w_{jl}w_{km}w_{kl} + 
w_{ik}w_{il}w_{jk}w_{lm}w_{jm} +w_{ik}w_{im}w_{jm}w_{jl}w_{kl} +
 w_{il}w_{im}w_{kl}w_{jk}w_{jm}].
\end{array}
\end{equation}
\par
To check the method we have first compared the results from the full sum of
permutations (1) with the results from the sum (3) cut at $K=4$ at 10
GeV. Both programs give the same results within a few permille
accuracy for all investigated distributions. 
Of special interest is the "BE ratio", defined for the pair of identical pions as 
a function of $Q = \sqrt {-(p_1 -p_2)^2}$
\begin{equation}
c_2(Q) = \frac {\int d^3p_1d^3p_2 \rho _2(p_1,p_2)\delta [Q-
\sqrt{-(p_1-p_2)^2}]} {\int d^3p_1d^3p_2 
\rho _1(p_1)\rho _1 (p_2)\delta [Q-
\sqrt{-(p_1-p_2)^2}]} \frac {<n>^2}{<n(n-1)>}.
\end{equation}
\par
Without weights it is rather flat and close to one, if we normalize separately
the numerator and the denominator of Eq. (5) to the same number of entries
(which is  achieved by the second factor  in (5)).  Including weights
produces a maximum at smallest $Q^2$  with the height about 2 (i.e. one unit
above the value at large $Q^2$) and a width $\sigma '$ about 0.15. Thus we
reproduce satisfactorily the shape assumed for the two-particle weight factor.
\par
At 30 GeV we do not have, as noted above, the results of full
symmetrization; for 174 events of the highest multiplicity the weights of the
previously generated events were attributed (still, this program requires 10 times more
computing time than the program with no more than 4 momenta symmetrized!). The
multiplicity distribution for two programs differs slightly in the tail,
although average multiplicities are quite similar: 10\% and 11\% higher than
without weights.
The peak at low $Q^2$ in the ratio of distributions exceeds slightly 2 and
looks similar in both programs.
\par
After this exercise we started generating events at {\bf  630 GeV}, the energy of the
{\sl UA1} experiment. For the calculations of weights we used the same value of $\sigma $
as before.
We did not have now the possibility   to estimate the results of
full symmetrization (the highest multiplicities of one sign pions exceeded 40 in
the $10^5$ events sample). Thus we have checked first that cutting the series (3)
at $K=3$ and at $K=4$  we get quite similar shapes of the $Q^2$
spectra, although the normalization is significantly different. Including
the term with $K=5$ we change even less all the  distributions. Thus we feel that 
cutting the series (3) at $K=5$ we 
get a reliable estimate of the results for $Q^2$  spectra from the weight method
(up to the possible change of normalization).
\par
The distribution of weights at $630 GeV$ is much broader than at previous
energies  and has a long tail (up to the values of a few
hundreds). Consequently, the multiplicity distribution is significantly changed
by the weighting.
Since the JETSET/PYTHIA parameters were fitted to reproduce inclusive
experimental data without weights, the change, e.g., of the average multiplicity induced by
weights should be compensated by the proper 	refitting procedures. Instead we have
applied  a simple method of multiplying weights by an extra $cV^n$
factor, where n is the number of  pions, and $c$ and $V$ are constants
fixed by the requirements to restore the original number of events and the
original average multiplicity. We return later to the details of this procedure. Such a rescaling 
of weights does not change significantly the shape
of $Q^2$ distributions. The BE ratio reflects mainly the assumed shape of the
two-particle weight (plus 1): for larger $\sigma$ it is wider and starts to increase 
above 2 for smallest  $Q^2$.
\par
The procedure seems to produce  too high a value of the BE ratio for
smallest  $Q^2$. As already noted, it is about twice the value for
small $Q^2$, whereas in most of the data it is only by some 50\% higher.
Let us note that we are using the old version of the  {\sl UA1} data [10] with rather
arbitrary normalization, and we attempt to describe them only down to the lower
limit of $Q^2$ about 0.01 $GeV^2$. In later publications of UA1 [13] the BE ratio
is shown to increase above 2 for lower $Q^2$. Since, however, this increase is still a subject of
controversy and anyway cannot be described by a Gaussian shape, we do not
discuss it here.

\par
To explain why the BE ratio does not increase up to the value of 2, one may invoke 
some coherent component, but a more obvious effect (which also lowers 
the BE ratio) is the existence of longer living resonances. Pions coming
from their decay are effectively "born" more than 10 fm from the collision
point. Thus the Gaussian width parameter in a two-particle weight for these
pions should by smaller by an order of magnitude, which allows practically to
neglect their contribution to be BE effect in the experimentally accessible  $Q^2$
range. Therefore the Bia{\l}as - Krzywicki weights should be calculated taking
into account only the permutations  of momenta of pions produced directly, or
resulting from the decay of the widest resonances.
\par
This is achieved easily if the procedure calculating weights is called before
the decay of long-living resonances, i.e. in the same place, where the original
LUBOEI procedure was called. We have rewritten correspondingly our program
separating the procedure LWBOEI (called directly from JETSET, and calculating
for each event a weight as a product of weights for positive, negative and
neutral "direct" pions) from the master program (calculating distributions with
and without weights). This procedure is available from authors as a FORTRAN
file. The results are now  (after rescaling the weights, as
desribed above) quite similar to the data  and may be brought to even
better agreement by fitting the only free parameter - the Gaussian
half-width of a two particle weight $\sigma$. We discuss this comparison with
data in more detail in the next section.  

\section{Results and comparison with data}
\par
We have generated $10^5$
events of $p \overline {p}$ minimum bias collisions at 630 GeV CM energy by the
default version of the PYTHIA/JETSET generator [6,11]. For
each event the weight factor was calculated by taking the 4-momenta of "direct"
pions of each sign, calculating for them a matrix of two - particle weights
$w_{ij}$ according to (2) with $\sigma  = 0.14$ GeV, and then the weight $w$ as a
series (3) cut at $K = 4$ or $5$. As already noted, the event weight is a product of
 weight factors for all three kinds of pions.
\par
Since the charged pion multiplicity distribution (used to fit the default values
 of the parameters with all weights equal 1) is strongly affected by weights, we
rescale the weight factors to restore the original average multiplicity. To this
end we multiply the weights by a $cV^n$ factor, where $n$ is the number of all
"direct" pions of the event, and $c$ and $V$ are calculated from the comparison
of the original and weighted multiplicity distribution. This is done by assuming
that the original multiplicity distribution of "direct" pions may be well
approximated by the negative binomial formula, i.e. that the NBD parameters
$\overline n$ and $1/k$ are given by the experimental values of $<n>$ and $<n(n-1>/<n>^2
-1$.
If with the weights we get a new average multiplicity $<n'>$, the original
value may then be restored by rescaling the weights with 
\begin{equation}
V = \frac{<n>(<n'>+k)}{<n'>(<n>+k)}
\end{equation}
 and 
\begin{equation}
c =\frac {[1+(1- V)<n'>/k]^k}{<w>},
\end{equation}
where $<w>$ is the average value of weights before
rescaling. We have checked that this procedure restores indeed the original
 average multiplicity with accuracy of few percent. On the other hand, the BE
ratios are little affected by rescaling (only the normalization, which is anyway
mainly a matter of convention, changes by a few percent).
\par
In Fig.1 we present the "BE ratio" (5) for pairs of positive
pions as a function of $x= ln_2(1GeV^2/Q^2)$ for the events from the original PYTHIA/JETSET generator
(without weights) and from our prescription with series (3) cut at $K=4$ and
$K=5$. In Fig.2 we show the "double ratio", i.e. the ratio of (5)  for pairs of
positive- and unlike sign pions for the same events.
\vspace{0.5cm}

\epsfxsize=9cm

~~~~~~~~~~~~~~~\epsfbox{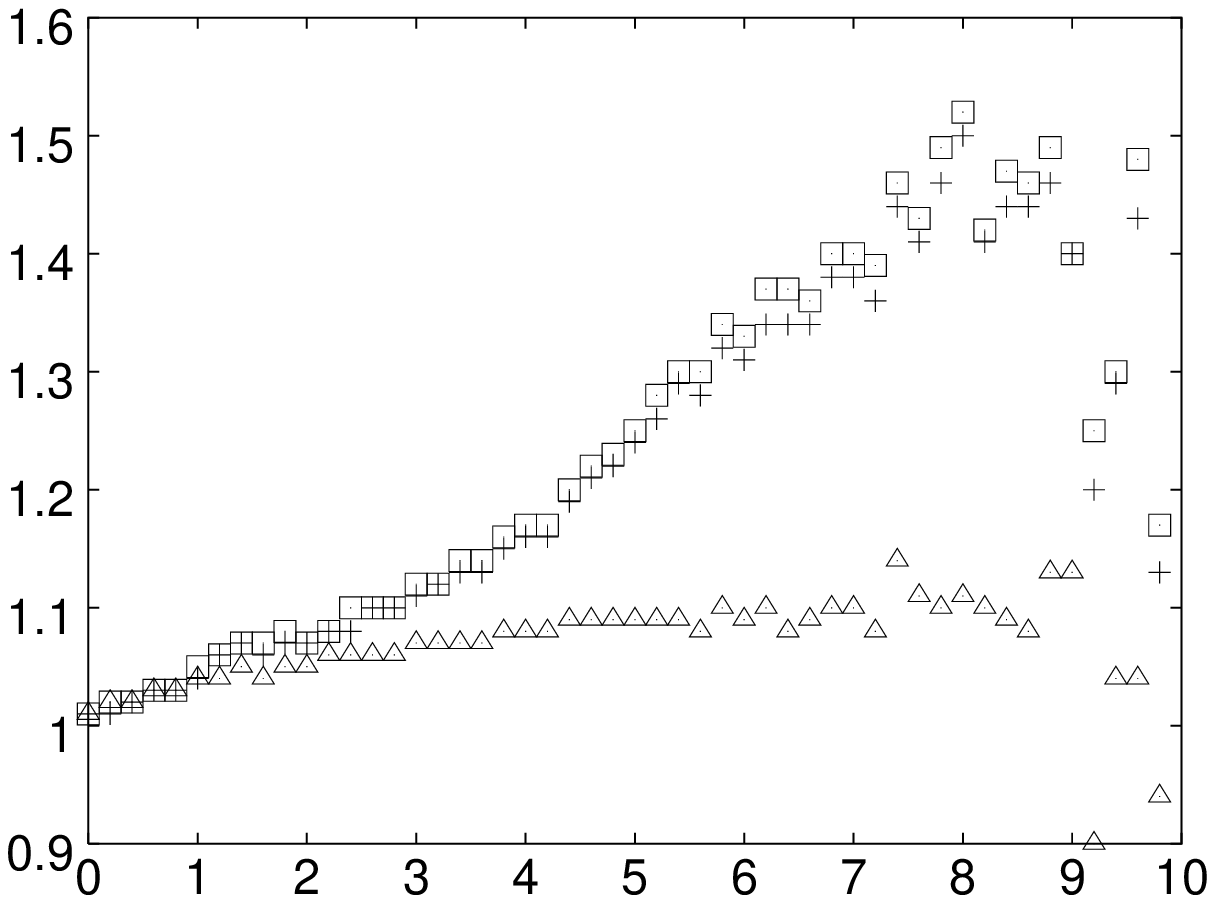}

\vspace{0.5cm}
\par
 {\bf Fig.1.} {\sl The   "BE ratio" (5) for positive pions as a
function of  $x= ln_2(1GeV^2/Q^2)$. Triangles, crosses and squares correspond to series (3) cut at
K   = 0 (no BE effect),\\ 4 and 5, respectively.}  

\vspace{0.5cm}

\epsfxsize=9cm

~~~~~~~~~~~~~~~\epsfbox{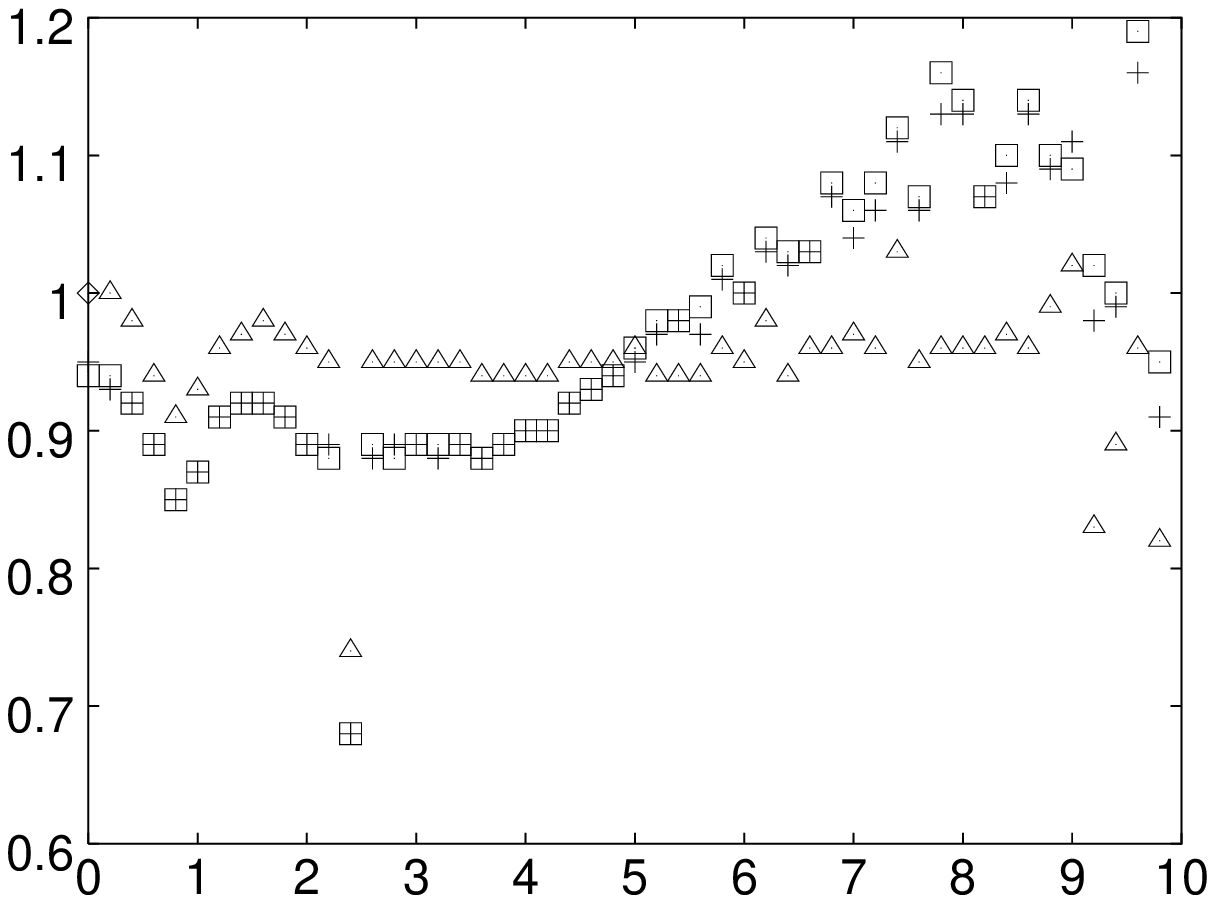}

\vspace{0.5cm}
\par
 {\bf Fig.2.} {\sl The   "double ratio" of ratios (5) for ++ and +-  pion pairs as a
function of  $x= ln_2(1GeV^2/Q^2)$. Triangles, crosses and squares correspond to series (3) cut
at\\
K = 0 (no BE effect), 4 and 5, respectively.}   \\

\par
We see that without weights both ratios are very close to one and depend weakly
on $Q^2$ (the dip in the double ratio at $x= 2.5, \ i.e. \ Q^2=0.17 GeV^2$ is the
reflection  of $K_s^0$ in unlike sign pairs, and the wider dip at lower x comes
from $\rho $). Our prescription produces a clear increase of both ratios at low
$Q^2$, and the difference between two choices of  $K_{max}$ are almost negligible.
Thus we believe that cutting the series (3) at $K = 5$ we approximate very well the results
with full formula for the weights (1), which would require an unreasonably long
computation time (even for supercomputers) when multiplicity exceeds 20.
\par
In Fig.3 and Fig.4 we compare our results obtained for $K \leq 5$ and two values of
$\sigma $ (0.14 and 0.1 GeV) with the {\sl UA1} data [10] normalized as in Eq.
(5).

\vspace{0.5cm}

\epsfxsize=9cm

~~~~~~~~~~~~~~~\epsfbox{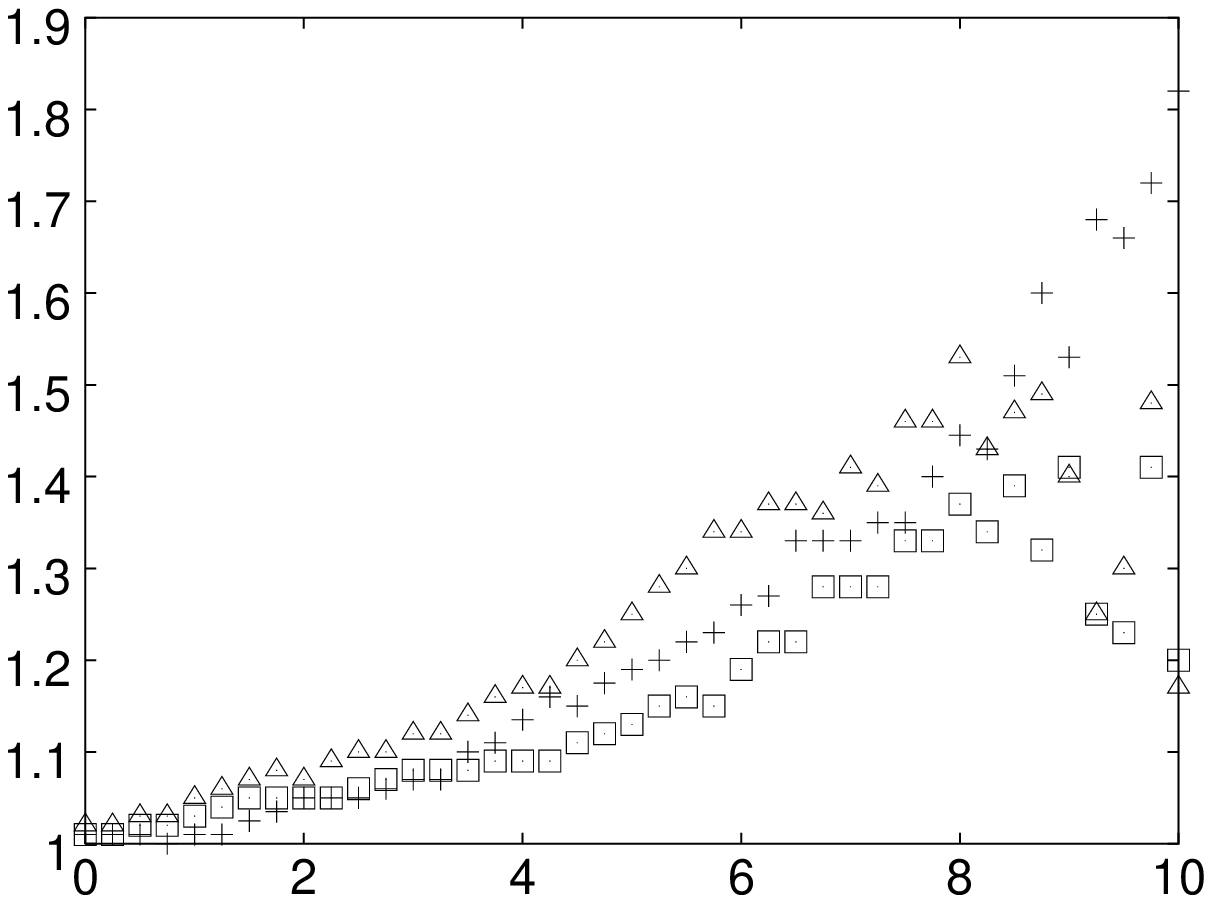}

\vspace{0.5cm}
\par
 {\bf Fig.3.} {\sl The   "BE ratio" (5) for positive pions as a
function of  $x= ln_2(1GeV^2/Q^2)$. Crosses represent  the {\sl UA1} data
[10],  triangles and squares correspond to the  series (3) cut at $K = 5$ with $\sigma = 0.14 \ GeV$ and $ \sigma = 0.1
\ GeV$, respectively.}   \\
\vspace{0.5cm}

\epsfxsize=9cm

~~~~~~~~~~~~~~~\epsfbox{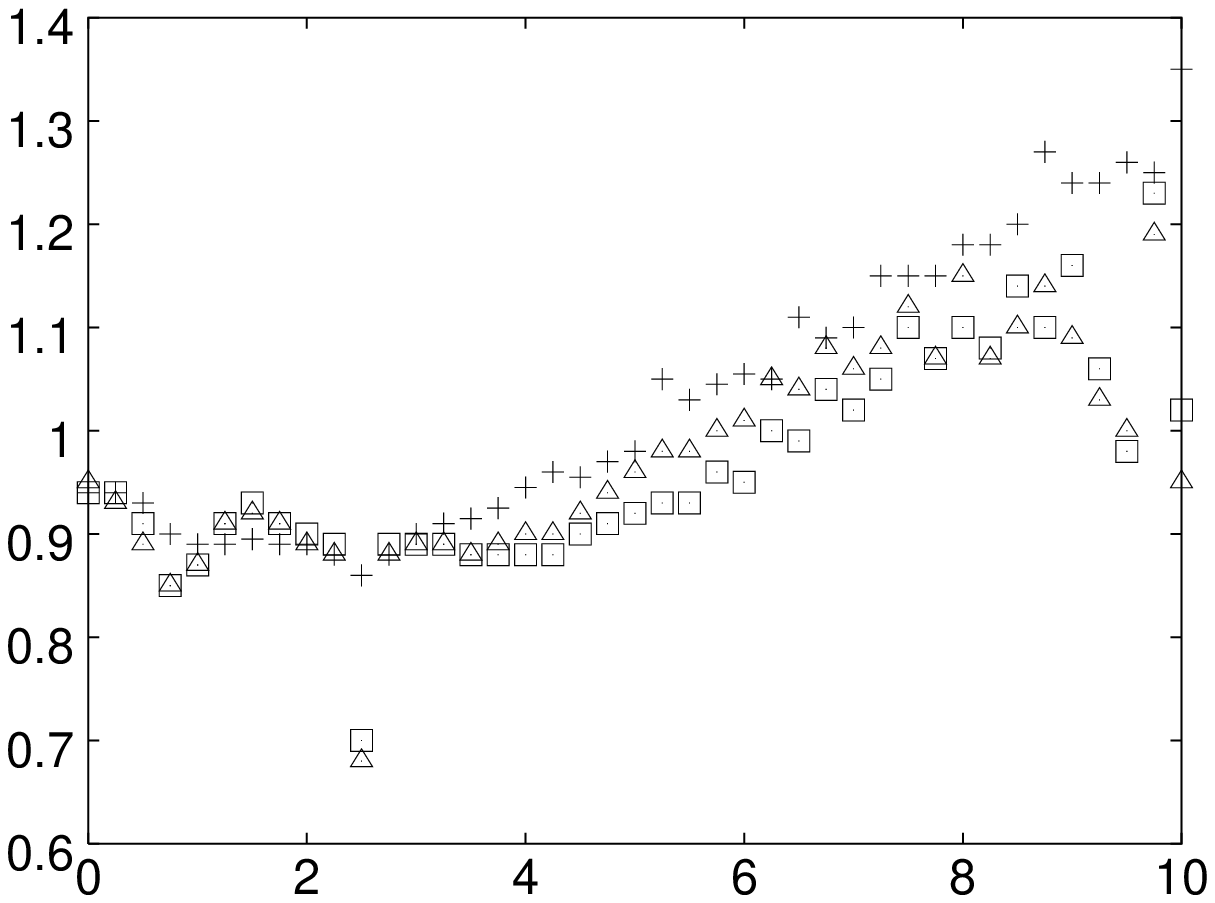}

\vspace{0.5cm}

\par
 {\bf Fig.4.} {\sl The   "double ratio" of ratios (5) for ++ and +-  pion pairs as a
function of  $x= ln_2(1GeV^2/Q^2)$. Crosses  represent  
the {\sl UA1} data [10], triangles and squares correspond 
to the  series (3) cut at $K = 5$ with $\sigma = 0.14 \ GeV$ and $ \sigma = 0.1
\ GeV$, respectively.}   \\

\par
 Let us stress here once
more that we use on purpose an old version of the data, which seemed to be well
described by a Gaussian shape of the ratio (5) for $Q^2>0.01 GeV^2 \\ (x  <  8)$ .
 Possible strong enhancement of
this ratio for lower values of $Q^2$ discussed in more detail in the later {\sl
UA1} papers [13], which seems to
signal a non-Gaussian shape, would require a modification of the shape of the
two-particle weight $w_{ij}$. Since here our purpose is merely to prove the
reliability of our method we do not want to enter this problem. In most of
the hadroproduction data  such low values of $Q^2$ are anyway not available 
 and the data seem to be relatively well fitted by a Gaussian.
\par
We see that the data for the BE ratio up to x = 8 are bracketed by the results
for two values of $\sigma $, corresponding to the source radius of 1 and 1.4 fm.
 We do not attempt to fit $\sigma $ here more precisely, since our purpose is
only to demonstrate the applicability of the method presented  above.
\section{Conclusions and outlook}
\par
We have shown that the weight method of implementing the Bose-Einstein interference
effects in Monte Carlo generators may be applied effectively to describe the
data. The prohibitive increase of computing time with multiplicity is avoided by
an approximation, in which only the selected class of terms ($K \leq K_{max}$ in
(3)) out of all $n!$ contributions is taken into account.
We show that already for $K_{max}=4$ and $K_{max}=5$  the results are almost the
same, which suggests that they approximate well those for the full series. 
\par
The change of multiplicity distributions induced by weights is compensated by
simple rescaling, equivalent to refitting of the original MC parameters. Using
the simple one parameter Gaussian form of two-particle weight factor we describe
reasonably well the {\sl UA1} data. 
\par 
There are many directions in which these results should be extended. One should
check if the weight method can describe higher order BE effects and
semi-inclusive ratios. The possibility of non-Gaussian and non-symmetric weight
factors should be investigated. Other hadroproduction processes should be
compared with $p \overline{p}$ collisions.   We hope to learn soon whether our
method is reliable enough to apply it confidently to the estimate of $W$ mass
shifts in the four-jet final states of the $e^+e^-\rightarrow W^+W^-$
collisions.\\  

\vspace{0.2cm}
{\large \bf Acknowledgements}
\vspace{0.2cm}
\par
We are grateful to A. Bia{\l}as and K. Zalewski for reading the manuscript and
useful remarks. A financial  support from KBN grants No 2 P03B 083 08 and No 2 P03B 196 09
is gratefully acknowledged. \\
\vspace{0.2cm}

{\large \bf References}
\vspace{0.2cm}         
\par
\noindent 1. S. Haywood, Rutherford Lab. Report RAL-94-074 (1995).
\par
\noindent 2. L. L\"onnblad and T. Sj\"ostrand, Phys. Lett. {\bf B351} (1995) 293.
\par
\noindent 3. K. Fia{\l}kowski and R. Wit,  Z. Phys. {\bf C}, in print
\par
\noindent 4. J. Ellis and K. Geiger,  {\sl Space, time and color in hadron production
via $e^+e^- \rightarrow Z^0$ and $e^+e^- \rightarrow W^+W^-$}, CERN preprint  CERN-TH.
95-283, October 1995, hep-ph/9511321.
\par
\noindent 5. S. Jadach and K. Zalewski, presented by S. Jadach at $Cracow
\  Epiphany\   Conference\\   on\   W \  Boson,\   1997$, to be published
in Acta Phys. Pol. {\bf B}.
\par
\noindent 6. T. Sj\"ostrand and M. Bengtsson, Comp.Phys.Comm. {\bf 43} (1987) 367;
T. Sj\"ostrand, CERN preprint CERN-TH.7112/93 (1993).
\par
\noindent 7. S. Pratt, Phys.Rev.Lett. {\bf 53} (1984) 1219.
\par
\noindent 8. A. Bia{\l}as and A. Krzywicki,  Phys. Lett. {\bf B354} (1995) 134.
\par
\noindent 9. J. Wosiek,  {\sl A simple formula for Bose-Einstein corrections},
preprint TPJU-1/97, January 1997, hep-ph/9701379.
\par
\noindent 10. F. Mandl and B. Buschbeck, {\sl Correlation integral studies in
DELPHI and in UA1}, in $Multiparticle \ Dynamics \ 1992$, C. Pajares (ed.), World
Sci. 1993. 
\par
\noindent 11. J. Pi\u{s}\'{u}t, N. Pi\u{s}\'{u}tov\'{a}, B. Tom\'{a}\u{s}ik,
Phys.Lett. {\bf B368} (1996) 179; Acta Phys. Slov. {\bf 46} (1996) 517.
\par
\noindent 12.T. Sj\"ostrand and M. Bengtsson, Comp.Phys.Comm. {\bf 46} (1987) 43.
\par
\noindent 13. N. Neumeister et  al. ({\sl UA1-Minimum  Bias  Coll.}),  
Z.Phys. {\bf C60} (1993) 633.
\end{document}